\documentclass[review]{elsarticle}

\usepackage{tikz-cd}
\usepackage{dsfont}
\usepackage{amsfonts}
\usepackage{amsmath}
\usepackage{slashed}
\usepackage{graphics}
\usepackage{amsmath}
\usepackage{amssymb}
\usepackage{tikz-cd}
\usepackage{slashed}
\usepackage{setspace}

\makeatletter
\def\ps@pprintTitle{%
 \let\@oddhead\@empty
 \let\@evenhead\@empty
 \def\@oddfoot{\centerline{\thepage}}%
 \let\@evenfoot\@oddfoot}
\makeatother

\journal{JHEAp - doi:10.1016/j.jheap.2018.01.001 }

\begin{document}

\begin{frontmatter}

\title{\textbf{Lorentz Invariance Violation effects on UHECR propagation: a geometrized approach}}

\tnotetext[t1]{Corresponding author email: marco.torri@unimi.it\\ \\ \\ JHEAp 18 (2018) 5-14 - doi:10.1016/j.jheap.2018.01.001}

\author[Unimi]{M.D.C. Torri*\corref{mycorrespondingauthor}}

\author[Unimi]{S. Bertini}

\author[Unimi]{M. Giammarchi}

\author[Unimi]{L. Miramonti}

\address[Unimi]{Dipartimento di Fisica\\
               Universit\'a degli Studi e INFN\\
               via Celoria 16, 20133, Milano, Italy}

\begin{abstract}
We explore the possibility to geometrize the interaction of massive fermions with the quantum structure of space-time, trying to create a theoretical background, in order to explain what some recent experimental results seem to implicate on the propagation of Ultra High Energy Cosmic Rays (UHECR). We will investigate part of the phenomenological implications of this approach on the predicted effect of the UHECR suppression, in fact recent evidences seem to involve the modification of the GZK cut-off phenomenon. The search for an effective theory, which can explain this physical effect, is based on Lorentz Invariance Violation (LIV), which is introduced via Modified Dispersion Relations (MDRs). Furthermore we illustrate that this perspective implies a more general geometry of space-time than the usual Riemannian one, indicating, for example, the opportunity to resort to Finsler theory.
\end{abstract}

\begin{keyword}
\emph{Lorentz invariance violation, UHECR, Finsler geometry, quantum gravity}
\end{keyword}

\end{frontmatter}

\section{Introduction}
Recent experimental observations, conducted on Ultra High Energy Cosmic Rays (UHECR), hint the possibility that the predicted Universe opacity to the propagation of this kind of highly energetic particles may be modified. Since the work of Coleman and Glashow \cite{Glash1}, many attempts of introducing LIV to justify such experimental evidences have been made, but this presents relevant difficulties, because currently there is no consistent theory of Quantum Physics and General Relativity. One of the greatest challenges in formulating such a unified theory is the impossibility of obtaining the energies needed to probe space-time at the Planck scale. In fact it is commonly believed that the Planck energy $E_{P}=\sqrt{\hbar c^5G^{-1}}$ and the Planck lenght $\lambda_{P}=\sqrt{G\hbar c^{-3}}$ represent the energy and the length scales separating the classical theory of gravity from the quantized one. Nevertheless, Planck-scale effects can possibly manifest themselves at lower energies as tiny violations of conservation laws. Several candidates quantum gravity theories have been proposed, such as loop quantum gravity, string theory, non commutative geometry, extensions to the Standard Model etc. These different models share the features of considering a modification of the dispersion relation (energy-momentum relation) of an elementary particle $E^2-p^2=m^2$ to the form $E^2-(1-f(p))p^2=m^2$ with $f(p)=\sum_{k=1}\alpha_{k}(E_{P})p^k$. Direct consequence of this modification is the violation of the Lorentz Invariance (LI) of the physical model studied. While LI as a global symmetry is a fundamental assumption of special relativity and the associated standard quantum field theory (standard model), it is not fundamental in General Relativity, in which the symmetry of space-time is given by classes of diffeomorphisms and LI is promoted from global to local symmetry as in a gauge theory. Global Lorentz invariance is an approximate symmetry that emerges in particular solutions of the Einstein field equations, in particular at low energies. So, even if there are no definitive evidences to sustain departures from LI, there are consistent hints indicating that \emph{Lorentz Invariance Violation} can be a theoretical consequence of quantum gravity.\\
To illustrate the scheme followed in this work, first we will introduce the concept of Modified Dispersion Relations (MDRs) as phenomena characterizing the quantum structure of space-time. Then we will obtain the geometry implied by the MDRs, introducing  Finsler structures.
We will illustrate some basic concepts of Finsler geometry to demonstrate the possibility to calculate the metric of the space-time starting from the metric of momentum (cotangent) field. After this we will introduce a minimal extension to the Standard Model, as an effective theory of the interaction of high energy particles with the quantum structure of space-time. In conclusion we will use this theory to determine the effects of quantum space on the propagation of Ultra-High Energy Cosmic Rays (UHECR), exploring some effects on their phenomenology.

\section{Lorentz Invariance Violation and Modified Dispersion Relation}

One simple way to introduce LIV consists in modifying the dispersion relations, considering this phenomenon caused by the quantum structure of gravity. In this work we assume, as reasonable physical hypothesis, that the dispersion relations are modified only for massive particles (leptons), while photons are considered Lorentz invariants. As illustrated in the work of Coleman and Glashow \cite{Glash1}, imposing a maximum speed, for a massive particle, lower to the speed of light, implies a modified dispersion relation, given by $E^2=(1-\epsilon)^2|\overrightarrow{p}|^2+(1-\epsilon)^4m^2$, which becomes, redefining the mass to reabsorb the correction term proportional to $m^2$:
\begin{equation}
\label{1}
E^2-(1-\epsilon)^2|\overrightarrow{p}|^2=m^2
\end{equation}
Generalizing the form of this MDR, introducing correction factors that depends on the instantaneous speed of the particles as in \cite{Smolin2}, it is possible to obtain:
\begin{equation}
\label{2}
f_{1}^{2}E^2-f_{2}^{2}p^2=m^2
\end{equation}
From this relation it is possible to obtain this explicit equality for the energy:
\begin{equation}
\label{3}
E=\sqrt{\frac{m^2}{f_{1}^{2}}+\frac{f_{2}^{2}}{f_{1}^{2}}\,p^2}\simeq pf_{3}\qquad where \;\;f_{3}=\frac{f_{2}}{f_{1}}
\end{equation}
Moreover we will consider the Modified Dispersion Relation (MDR) type introduced in Liberati et al. work \cite{Liberati1}, which has the form:
\begin{equation}
\label{7}
E^2-(1-f(|\overrightarrow{p}|))|\overrightarrow{p}|^2=m^2
\end{equation}
with $f(|\overrightarrow{p}|)$ that is a tiny perturbation, depending only on the magnitude of the three-momentum $\overrightarrow{p}$. It can be demonstrated that a consistent category of MDRs can be written in this particular form, those that preserve, in some measure, the space isotropy.
It is important to stress that this type of MDRs require the reintroduction of the concept of \emph{privileged frame of reference}.\\
In special relativity, one can use the ordinary dispersion relation to obtain the metric of the momentum space (the Minkowski case) that coincides with the metric of the coordinates space. In fact the ordinary dispersion relation can be written as:
\begin{equation}
\label{8}
E^2-\overrightarrow{p}^2=m^2\Rightarrow \eta^{\mu\nu}p_{\mu}p_{\nu}=m^2
\end{equation}
In order to preserve the metric derivation in momentum space of the MDR, we decide to resort to homogeneous perturbations, so, assuming their analyticity, the form of the perturbation functions must be:
\begin{equation}
\label{9}
f\left(\frac{|\overrightarrow{p}|}{E}\right)=\sum_{k=1}^{\infty}\alpha_{k}\left(\frac{|\overrightarrow{p}|}{E}\right)^k
\end{equation}
instead of the usual form:
\begin{equation}
\label{10}
f(|\overrightarrow{p})=\sum_{k=1}^{\infty}\beta_{k}\left(\frac{|\overrightarrow{p}|}{M_{p}}\right)^k
\end{equation}
with $M_{p}$ that represents the Planck mass, a fixed scale that suppresses the perturbation magnitude.\\
In this way eq. (\ref{7}) becomes:
\begin{equation}
\label{11}
E^2-\left(1-f\left(\frac{|\overrightarrow{p}|}{E}\right)\right)|\overrightarrow{p}|^2=m^2
\end{equation}
It must be highlighted the importance of being cautious in defining the perturbation function (\ref{9}), in order to guarantee the existence of real solution for the energy in the previous equation. This can be obtained posing appropriate constrains on the coefficients of the series. In this way it can be showed that $E$ assumes positive finite values, as function of $p$, and:
\begin{equation}
\label{12}
\lim_{p\rightarrow\infty}{\frac{|\overrightarrow{p}|}{E}}=1+\delta
\end{equation}
for a tiny positive constant $\delta$, reobtaining the Coleman and Glashow scenario, if the perturbation function admits the limit $f(1+\delta)=\epsilon$. So using homogeneous perturbation functions is of great interest because hence it is possible to construct a continuous transition from the classic GZK effect to the Coleman and Glashow foreseen suppression. This permits to investigate the possibility of a dilatation of the GZK opacity sphere radius till an infinite limit. An explicit example of homogeneous perturbations is given in \cite{Koste4}, where this kind of modifications is obtained as a special case of a general Lorentz violating hermitian quantum field theory. In the cited work hermitian perturbing hamiltonians imply that the energy shift is real.\\
Now it is possible to find out the "\emph{personal maximum speed}" of the particle, using the Hamilton-Jacobi equation for the energy (\ref{3}):
\begin{equation}
\begin{split}
\label{4}
&\frac{\partial}{\partial p}E=c=f+\left(-\frac{p}{E^2}\right)f'\partial_{p}E+f'\frac{1}{E}\;\Rightarrow\\
\Rightarrow\;&\frac{\partial}{\partial p}E=f+\left(\frac{1}{E}-\frac{p}{E^2}\right)f'\simeq 0
\end{split}
\end{equation}
in the limit of great magnitude for $E$ and $\overrightarrow{p}$. Finally from this relation:
\begin{equation}
\label{5}
\partial_{p}E=f
\end{equation}
and the modified dispersion relation therefore can be written in the form:
\begin{equation}
\label{6}
E^2-f^2p^{2}=m^2
\end{equation}
where $f$ represents the modified maximum attainable velocity for every given three-momentum of the particle. In this sense a propagating particle feels a local space-time (tangent space) foliation, that is parameterized by the particle momentum. This means that every particle lives in a space-time defined by the magnitude of its velocity. From this the necessity to resort to a geometrical structure, which can deal with this dependence, that is the Finsler geometry.\\
Following this approach, MDR (\ref{7}) can be written as:
\begin{equation}
\label{13}
\widetilde{g}(p)^{\mu\nu}p_{\mu}p_{\nu}=F^2(p)=m^2
\end{equation}
with the metric for (\ref{10}) given by:
\begin{equation}
\label{14}
\widetilde{g}(p)^{\mu\nu}=\left(
                         \begin{array}{cccc}
                           1 & 0 & 0 & 0 \\
                           0 & -(1-f(p)) & 0 & 0 \\
                           0 & 0 & -(1-f(p)) & 0 \\
                           0 & 0 & 0 & -(1-f(p)) \\
                         \end{array}
                       \right)
\end{equation}
The metric is calculated via the relation:
\begin{equation}
\label{15}
\widetilde{g}^{\mu\nu}(p)=\frac{1}{2}\frac{\partial}{\partial p_{\mu}}\frac{\partial}{\partial p_{\nu}}F^{2}(E,\,\overrightarrow{p})
\end{equation}
where $F^2(E,\,\overrightarrow{p})$ represents the MDR and $F(E,\,\overrightarrow{p})$ is a 1 degree homogeneous positive definite Finsler norm.\\
The metric obtained is reported to a diagonal form, eliminating a non-diagonal part, which gives a null contribution in the computation of the value of the MDR.\footnote{It is possible to demonstrate that if the correction function $f$ is homogeneous of 0 order the non-diagonal part of the metric gives null contribution in computing the dispersion relation, so it is a reasonable choice to eliminate it from the obtained metric.} This is in agreement with the fact that every massive particle has a \emph{personal maximum attainable velocity}. Furthermore the previous metric is defined on the momenta space and must be converted to a metric defined on the coordinates space, using the Legendre transformation, but before it is necessary to introduce the basics of Finsler geometry, as already hypothesized in \cite{Koste3}.

\section{Finsler geometry}
Following the classical way to introduce Finsler Geometry, as in \cite{Liberati1}, we can define a Finsler manifold as a geometric structure $M$, where in each tangent space a norm is defined, which is not necessarily induced by an inner product. The norm is a real function of the section of the tangent space $T_{x}M$, so depends on the point $x$ and on a vector $v\in T_{x}M$. This function must satisfy the usual definition of \emph{norm}, so it must be positive definite and 1-degree homogeneous:\begin{itemize}
                          \item $F(x,\,v)>0$  $\forall v\neq0,\,x\in M$
                          \item $F(x,\,\lambda v)=|\lambda|F(x,\,v)$
                        \end{itemize}
Contrary to the Riemannian case, here it is the norm that induces an inner product, parametrized not just by points belonging to the variety $M$, but also by vectors in $T_{x}M$. So, in Finsler geometry one generalizes the Riemaniann metric by using  the squared norm to obtain:
\begin{equation}
\label{16}
g_{\mu\nu}(x,\,v)=\frac{1}{2}\frac{\partial^2F(x,\,v)^2}{\partial v^{\mu}\partial v^{\nu}}
\end{equation}
which is required to have $\det{g_{\mu\nu}}\neq0$ and $g_{\mu\nu}\in\mathbb{C}$.\\
In this way it is possible to reobtain the norm from the inner product in the common form:
\begin{equation}
\label{17}
F(x,\,v)=\sqrt{g(x,\,v)_{\mu\nu}v^{\mu}v^{\nu}}
\end{equation}
The fact that the metric $g(x,\,v)_{\mu\nu}$ is not positively defined implicates that we are not dealing with a real Finsler structure, but with a so called pseudo-Finslerian manifold \cite{Koste1}. A final remark on the fact that all the results about Finsler geometry present in this work remain valid.\\
As in Riemann geometry, even in this case it is possible to define duality between vectors and dual forms, using the metric:
\begin{equation}
\label{18}
\omega_{\mu}=g_{\mu\nu}(x,\,v)v^{\nu}
\end{equation}
Now we underline the main result of this section, useful for the prosecution of this work, about the Legendre transformation of the Finsler metric.
First of all we have to introduce the concept of dual Finsler norm, defined as:
\begin{equation}
\label{19}
F^{*}(\xi)=\max{\{\xi(y):y\in V,F(y)=1\}},\quad \xi\in V^{*}
\end{equation}
that is well posed and finite, because the set $\{y\in V:F(y)=1\}$ is compact.\\
The Legendre transformation is a function $l:V\rightarrow V^{*}$ defined via the relation:
\begin{equation}
\label{20}
l(y)_{\nu}=g(y)_{\mu\nu}y^{\mu}
\end{equation}
What is possible to demonstrate now is a result\footnote{The proof of this statement is given in appendix.}, useful for the proceeding of this work:\\
\emph{Proposition}:
\begin{enumerate}
  \item $F=F^{*}\circ l$
  \item \emph{the Legendre transformation is a bijection}.
\end{enumerate}
Using the biunivocal correspondence between the metric and its Legendre transformation, it is possible to consider a Finsler space as $T\,TM$, with $g_{\mu\nu}$ defined and acting on the variety $TM$.

\section{Legendre transformation}
To redefine the previous metric (\ref{14}), to eliminate its dependence from quantities defined on the cotangent space, it is necessary to resort to the Legendre transformation. Let us start by defining the function:
\begin{equation}
\label{21}
\dot{x}^{\mu}=\frac{1}{2}\left(\frac{\partial}{\partial p_{\mu}}F^2\right)
\end{equation}
as the classical velocity associated to the momentum of a particle. Now it is simple to demonstrate that:
\begin{equation}
\label{22}
p_{\mu}\dot{x}^{\mu}=F^2
\end{equation}
in fact, due to the homogeneity of order 2 of the function $F^2$, it follows:
\begin{equation}
\label{23}
p_{\mu}\dot{x}^{\mu}=p_{\mu}\left(\frac{1}{2}\frac{\partial F^2}{\partial p_{\mu}}\right)=F^2
\end{equation}
Furthermore, using the $\widetilde{g}_{\mu\nu}$ metric, associated with $g_{\mu\nu}$ via the Legendre transformation, it is easy to proof that:
\begin{equation}
\label{24}
\widetilde{g}^{\mu\nu}p_{\nu}=\dot{x}^{\mu}=\frac{1}{2}\left(\frac{\partial^2 F^2}{\partial p_{\mu}\partial p_{\nu}}p_{\nu}\right)=\frac{\partial}{\partial p_{\nu}}\left(\frac{1}{2}\frac{\partial}{\partial p_{\mu}}F^2\right)p_{\nu}
\end{equation}
using the fact that the function:
\begin{equation}
\label{25}
\frac{1}{2}\left(\frac{\partial}{\partial p_{\mu}} F^2\right)
\end{equation}
is homogeneous of degree 1.\\
From equation (\ref{24}) and the last proposition of the previous section, it follows that the function $\dot{x}^{\mu}(p)$ is univocally defined. This means that the Legendre transformation of the metric $\widetilde{g}_{\mu\nu}(p)$, given by $g_{\mu\nu}(\dot{x})$, is well defined and the two are in biunivocal correspondence. From these statements it follows:
\begin{equation}
\label{26}
m^2g_{\mu\nu}\dot{x}^{\mu}\dot{x}^{\nu}=g_{\mu\nu}\widetilde{g}^{\mu\alpha}\widetilde{g}^{\nu\beta}p_{\alpha}p_{\beta}=F^2=m^2
\end{equation}
Moreover the obtained metric depends not only on the point that defines the local tangent space, but even on a vector, as in case of Finsler space.
Finally it is possible to write the metric of the tangent space in the form:
\begin{equation}
\label{27}
g_{\mu\nu}=\widetilde{g}_{\mu\nu}=\left(
                                       \begin{array}{cccc}
                                          1 & 0 & 0 & 0 \\
                                          0 & -(1+f(p)) & 0 & 0 \\
                                          0 & 0 & -(1+f(p)) & 0 \\
                                          0 & 0 & 0 & -(1+f(p)) \\
                                       \end{array}
                                    \right)
\end{equation}

\section{Vierbein and induced symmetries}
In the previous sections we have illustrated that MDRs can be integrated in a theory, which implies that every massive particle feels a space-time characterized by a family of metrics, parameterized by the velocity of the particle itself \cite{Smolin1}. This generates a family of parameterized orthonormal frame fields, the \emph{vierbein} or \emph{tetrad}, which allow to write the metric as:
\begin{equation}
\label{28}
g_{\mu\nu}=\eta_{ab}\,\left[e\right]^{\,a}_{\mu}(p)\,\left[e\right]^{\,b}_{\nu}(p)
\end{equation}
where the latin indices represent the global coordinates and the greek ones the local coordinates defined on the tangent space.\\
The explicit form of the \emph{vierbein}, in case of metric (\ref{14}), is given by:
\begin{equation}
\label{29}
\left[e\right]_{\mu}^{\,a}=\left(
                             \begin{array}{cc}
                               1 & \overrightarrow{0} \\
                               \overrightarrow{0}^{t} & \sqrt{1+f(p)}\,\mathbb{I} \\
                             \end{array}
                           \right)
\end{equation}
Using the \emph{tetrad} it is possible to define the local form for every quantity of interest, such as the Clifford Algebra, generated by the Dirac gamma matrices:
\begin{equation}
\label{30}
\Gamma_{\mu}=\left[e\right]_{\mu}^{\,a}\gamma_{a}
\end{equation}
and the Lorentz group:
\begin{equation}
\label{31}
\Lambda(x)^{\mu}_{\,\nu}=\left[e\right]^{\mu}_{\,a}\Lambda^{a}_{\,b}\left[e\right]^{\,b}_{\nu}
\end{equation}
obtaining a non-linear realization of the Lorentz group, that preserves the form of the MDR, result comparable to \cite{Smolin1}. The action of the obtained group on the 4-vector $p^{\mu}=(E,\,\overrightarrow{p})$, preserves even the homogeneity of degree 0 of the correction functions $f$ (see Appendix B). The possibility to use the modified Lorentz group to transform the coordinates from one inertial frame to another, associated with the fact that MDRs considered in this work are generated from a metric, imply that the form (\ref{11}) remains constant in every inertial system.\\
As explained in \cite{Liberati1}, every massive particle feels a different Finsler structure, governing the dynamic of the particle itself, interacting with the quantum structure of the space-time. The following scheme illustrates how to use the \emph{vierbein} to project vectors from a tangent space with the metric $g_{\mu\nu}(x,\,v)$ to a space where it is defined another metric $\overline{g}(y,\,w)_{\mu\nu}$, this will be useful in computing the free energy for reactions involving particles with different masses.
\[\begin{tikzcd}
         (TM,\,\eta_{ab},\,v) \arrow{d}{\left[e\right]} \arrow{rr}[swap]{\Lambda} && (TM,\,\eta_{ab},\,w)\arrow{d}[swap]{\left[\overline{e}\right]} \\
        (T_{x}M,\,g_{\mu\nu}(x,\,v)) \arrow{rr}[swap]{\left[\overline{e}\right]\circ\Lambda\circ\left[e^{-1}\right]} && (T_{x}M,\,\overline{g}_{\mu\nu}(y,\,w))
\end{tikzcd}\]

\section{Modified connection and redefined Dirac equation}
We can now outline the features of the Dirac equation for the "\emph{Finslerian curved}" space-time, whit which we are dealing. Starting from the generic form of the Dirac equation for curved space-time:
\begin{equation}
\label{32}
\left(i\gamma^{a}\left[e_{\,a}^{\mu}\right]D_{\mu}-m\right)\psi=0\Rightarrow \left(i\Gamma^{\mu}D_{\mu}-m\right)\psi=0
\end{equation}
where $D_{\mu}$ is the total covariant derivative acting on a tensor with two indices, one local (greek) and the other global (latin). It is related to the Christoffel and the Cartan connections as follows:
\begin{equation}
\label{33}
D_{\mu}v^{\nu}_{\,a}=\partial_{\mu}v^{\nu}_{\,a}+\Gamma_{\mu\alpha}^{\,\nu}v^{\alpha}_{\,a}-\omega_{\mu b}^{\,a}v^{\nu}_{\,b}
\end{equation}
The Christoffel connection is defined, as usual, as:
\begin{equation}
\label{34}
\Gamma_{\mu\nu}^{\,\alpha}=\frac{1}{2}g^{\alpha\beta}\left(\partial_{\mu}g_{\beta\nu}+\partial_{\nu}g_{\mu\beta}-\partial_{\beta}g_{\mu\nu}\right)
\end{equation}
Using the metric obtained by the Legendre transformation (\ref{27}), it is possible to find out that the not null components of the Christoffel connection are proportional to:
\begin{equation}
\label{35}
\frac{\partial}{\partial x^{\mu}}g_{\alpha\beta}=-\frac{\partial}{\partial x^{\mu}}f(p(x))=-\frac{\partial}{\partial p}f(p)\frac{\partial}{\partial x^{\mu}}p(x)
\end{equation}
Every term turns out to be small ($|\partial_{x_{\mu}}p(x)|\ll1$) because the geometrized interaction of a massive particle with the quantum structure of space-time is small, even at high energies, and $|\partial_{p}f(p)|\ll1$ because of the form of the perturbation function (\ref{9}). In fact for example, at high energies:
\begin{equation}
\label{36}
\begin{split}
&\partial_{p^{j}}f(p)=\partial_{p^{j}}\sum_{k}\alpha_{k}\frac{|\overrightarrow{p}|^{k}}{E^{k}}=\partial_{p^{j}}\sum_{k}\alpha_{k}\frac{|\overrightarrow{p}|^{k}}{(\sqrt{|\overrightarrow{p}|^2+m^2})^k}=\\
=&\sum_{k}\left(\alpha_{k}k\frac{|\overrightarrow{p}|^{k-2}p_{j}}{(\sqrt{|\overrightarrow{p}|^2+m^2})^k}-\alpha_{k}k\frac{|\overrightarrow{p}|^{k}p_{j}}{(\sqrt{|\overrightarrow{p}|^2+m^2})^{k+2}}\right)\rightarrow0
\end{split}
\end{equation}
where it has been used the equivalence $E\simeq\sqrt{|\overrightarrow{p}|^2+m^2}$, so we can write the Christoffel symbols as:
\begin{equation}
\label{37}
\begin{split}
&\Gamma_{\mu0}^{\,0}=\Gamma_{00}^{\,i}=\Gamma_{\mu\nu}^{\,i}=0\qquad\forall \mu\neq\nu\\
&\Gamma_{ii}^{\,0}=-\frac{1}{2}\partial_{0}f(p)\simeq0\\
&\Gamma_{0i}^{\,0}=\Gamma_{i0}^{\,0}=\frac{1}{2(1-f(p))}\partial_{0}f(p)\simeq0\\
&\Gamma_{ii}^{\,i}=\frac{1}{2(1-f(p)-g(p))}\partial_{i}f(p)\simeq0\\
&\Gamma_{jj}^{\,i}=-\frac{1}{2(1-f(p))}\partial_{i}f(p)\simeq0\qquad\forall i\neq j\\
&\Gamma_{ij}^{\,i}=\Gamma_{ji}^{\,i}=\frac{1}{2(1-f(p))}\partial_{i}f(p)\simeq0\qquad\forall i\neq j
\end{split}
\end{equation}
where the latin indices belong to the set $\{1,\,2,\,3\}$ and the greek ones to $\{0,\,1,\,2,\,3\}$.\\
Whit the Christoffel connection it is possible to define the local covariant derivative, which acts on vectors in the following manner:
\begin{equation}
\label{38}
\nabla_{\mu}v^{\nu}=\partial_{\mu}v^{\nu}+\Gamma_{\mu\alpha}^{\,\nu}v^{\alpha}\simeq\partial_{\mu}v^{\nu}
\end{equation}
showing that we can replace it with the standard partial derivative. Using the local covariant derivative (\ref{37}) we can define the Cartan connection in the following way:
\begin{equation}
\label{39}
\omega_{\mu ab}=\left[e\right]^{\nu}_{\,a}\nabla_{\mu}\left[e\right]_{\nu b}\simeq\left[e\right]^{\nu}_{\,a}\partial_{\mu}\left[e\right]_{\nu b}
\end{equation}
We set now the external forms:
\begin{equation}
\label{40}
\begin{split}
&e_{0}^{\,\mu}=dx^{\mu}\\
&e_{i}^{\,\mu}=\sqrt{1-f(p)}dx^{\mu}
\end{split}
\end{equation}
Using the first Cartan structural equation:
\begin{equation}
\label{41}
de=e\wedge\omega
\end{equation}
it is possible to obtain the following expression, valid for the non-zero components of the spinorial connection:
\begin{equation}
\label{42}
\begin{split}
&\frac{1}{2}\frac{-1}{\sqrt{1-f}}\;\epsilon_{ijk}\left(\partial_{x^{i}}fdx^{j}\wedge dx^{i}+\partial_{x^{k}}fdx^{k}\wedge dx^{i}\right)=\\
&=\sqrt{1-f}\;\epsilon_{ijk}\left(dx^{j}\wedge\omega^{ij}+dx^{k}\wedge\omega^{ik}\right)
\end{split}
\end{equation}
From relation (\ref{36}), it follows that the only components of the connection differential forms $\omega$, that are not identically equal to zero, are:
\begin{equation}
\label{43}
\begin{split}
&\omega^{12}=-\frac{1}{2}\frac{1}{1-f}\left(\partial_{y}fdx-\partial_{x}fdy\right)\\
&\omega^{13}=-\frac{1}{2}\frac{1}{1-f}\left(\partial_{z}fdx-\partial_{x}fdz\right)\\
&\omega^{23}=-\frac{1}{2}\frac{1}{1-f}\left(\partial_{z}fdy-\partial_{y}fdz\right)\\
\end{split}
\end{equation}
So even for the Cartan connection the non-zero coefficients are proportional to elements like (\ref{35}). From this follows that: $\omega_{\mu ab}\simeq 0$ for every choice of indices and the total covariant derivative can be approximated by the standard derivative $D_{\mu}\simeq \partial_{\mu}$. In conclusion, the solution obtained for the connection induced by the geometrization of the interaction of a massive particle with the quantum structure of space-time implies that the Finslerian structure of the space-time is asymptotically flat.\\
To obtain the required modification of the Dirac equation it is necessary to introduce now the modified gamma matrices, present in (\ref{32}), which are those defined in relation (\ref{30}). They can be obtained satisfying the Clifford Algebra definitory relation:
\begin{equation}
\label{44}
\{\Gamma_{\mu},\Gamma_{\nu}\}=2g_{\mu\nu}=2\left[e\right]_{\mu}^{\,a}\eta_{ab}\left[e\right]_{\nu}^{\,b}
\end{equation}
Starting from the Dirac representation of the $gamma$ matrices, we obtain the modified ones:
\begin{equation}
\label{45}
\begin{split}
&\Gamma_{0}=\gamma_{0}\qquad \Gamma_{i}=\sqrt{1+f(p(x,\,\dot{x}))}\;\gamma_{i}\\
&\Gamma^{0}=\gamma^{0}\qquad \Gamma^{i}=\frac{1}{\sqrt{1+f(p(x,\,\dot{x}))}}\;\gamma^{i}
\end{split}
\end{equation}
The $\Gamma_5$ matrix can be defined using the relation:
\begin{equation}
\label{46}
\Gamma_5=\frac{\epsilon^{\mu\nu\alpha\beta}}{4!}\Gamma_{\mu}\Gamma_{\nu}\Gamma_{\alpha}\Gamma_{\beta}=\frac{1}{\sqrt{\det{g}}}\Gamma_{0}\Gamma_{1}\Gamma_{2}\Gamma_{3}=\gamma_{5}
\end{equation}
where the curved spacetime total antisymmetric tensor has been used.\\
The Dirac equation (\ref{32}) assumes the form, as in \cite{Koste2}:
\begin{equation}
\label{47}
\left(i\Gamma^{\mu}\partial_{\mu}-m\right)\psi=0
\end{equation}
Now it is necessary to introduce the modified spinors, so following the classical argumentation to obtain the general ones, we start from the representation of the spinors for null three-momentum.
Starting from the solution of the Dirac equation expressed, using plane waves, in the form:
\begin{equation}
\label{48}
\begin{split}
&\psi^{+}(x)=u_{r}(p)e^{-ip_{\mu}x^{\mu}}\\
&\psi^{-}(x)=v_{r}(p)e^{ip_{\mu}x^{\mu}}
\end{split}
\end{equation}
and considering only the positive energy spinor defined for null three-momentum $\overrightarrow{p}=0$:
\begin{equation}
\label{49}
u_{r}(m,\,\overrightarrow{0})=\chi_{r}=\left(
                                         \begin{array}{c}
                                           1 \\
                                           0 \\
                                         \end{array}
                                       \right)
\end{equation}
it is possible to obtain the generic one, starting from the fact that the modified Dirac equation implies:
\begin{equation}
\label{50}
(i\Gamma^{\mu}\partial_{\mu}-m)u_{r}(p)e^{-ip_{\mu}x^{\mu}}\Rightarrow(\slashed{p}-m)u_{r}(p)=0
\end{equation}
and that it is true the relation:
\begin{equation}
\label{51}
(\slashed{p}-m)(\slashed{p}+m)=(p^{\mu} p_{\mu})-m^2=0\;\Rightarrow \;(\slashed{p}-m)(\slashed{p}+m)u_{r}(m,\,\overrightarrow{0})=0
\end{equation}
Now it is possible to obtain the general spinor by:
\begin{equation}
\label{52}
\begin{split}
&(\Gamma^{\mu}p_{\mu}+m)\left(
                          \begin{array}{c}
                            \chi_{r} \\
                            0 \\
                          \end{array}
                        \right)
\Rightarrow\\
\Rightarrow&
\left(p_{0}\left(
             \begin{array}{cc}
               \mathbb{I} & 0 \\
               0 & \mathbb{-I} \\
             \end{array}
           \right)-
p_{i}\left(
       \begin{array}{cc}
         0 & -\sigma^{i} \\
         \sigma^{i} & 0 \\
       \end{array}
     \right)
\sqrt{1-f}\right)
\left(
  \begin{array}{c}
    \chi_{r} \\
    0 \\
  \end{array}
\right)+\\
+&m\left(
   \begin{array}{cc}
     \mathbb{I} & 0 \\
     0 & \mathbb{I} \\
   \end{array}
 \right)
 \left(
   \begin{array}{c}
     \chi_{r} \\
     0 \\
   \end{array}
 \right)
 =\left(
     \begin{array}{c}
       (E+m)\chi_{r} \\
       \overrightarrow{p}\overrightarrow{\sigma}\sqrt{1-f}\;\chi_{r} \\
     \end{array}
   \right)
\end{split}
\end{equation}
where the standard representation for the Pauli $\sigma$ matrices has been used.\\
The normalization of the states created can be obtained by:
\begin{equation}
\label{53}
\begin{split}
&\overline{u}_{r}(m,\,\overrightarrow{0})(\slashed{p}+m)^2u_{s}(m,\,\overrightarrow{0})=\overline{u}_{r}(m,\,\overrightarrow{0})(p^{\mu}p_{\mu}+2\slashed{p}m+m^2)u_{s}(m,\,\overrightarrow{0})=\\
=&\overline{u}_{r}(m,\,\overrightarrow{0})(2m(\slashed{p}+m))u_{s}(m,\,\overrightarrow{0})=2m(E+m)\delta_{rs}
\end{split}
\end{equation}
so the final form for a generic positive-energy spinor is given by:
\begin{equation}
\label{54}
u_{r}(p)=\frac{1}{\sqrt{2m(E+m)}}\left(
                                   \begin{array}{c}
                                     (E+m)\chi_{r} \\
                                     \overrightarrow{p}\overrightarrow{\sigma}\sqrt{1-f}\;\chi_{r}\\
                                   \end{array}
                                 \right)
\end{equation}
In analogy, one can easily obtain the generic negative energy spinor.\\
Having defined the modified spinors from the plane-waves expansion of the Dirac equation in the (\ref{27}) metric, it is now possible to verify its compatibility with the MDR. In fact:
\begin{equation}
\label{55}
\begin{split}
&\left(i\Gamma^{\mu}\partial_{\mu}+m\right)\left(i\Gamma^{\nu}\partial_{\nu}-m\right)u(p)e^{ip_{\mu}x^{\mu}}\Rightarrow\\
\Rightarrow&\left(\frac{1}{2}\{\Gamma^{\mu},\Gamma^{\nu}\}p_{\mu}p_{\nu}-m^2\right)u(p)=0\Rightarrow\\
\Rightarrow&\left(p_{\mu}p_{\nu}g^{\mu\nu}-m^2\right)u(p)=0\Rightarrow\\
\Rightarrow& E^2-|\overrightarrow{p}|^2(1-f(p))-m^2=0
\end{split}
\end{equation}

\section{Standard Model extension}
Now with the redefined Dirac equation it is possible to modify the Standard Model, obtaining an effective theory, the SM minimal extension, useful to compute the interaction of a UHECR proton with the CMB. As an example we consider what happens to QED, therefore, it is so possible to define the effective Lagrangian for a free electron:
\begin{equation}
\label{56}
\overline{\psi}\left(i\Gamma^{\mu}\partial_{\mu}-m\right)\psi
\end{equation}
This term of the Lagrangian is defined in the modified tangent space $(T_{x}M,\,g_{\mu\nu})$.\\
Considering the modified gamma matrices it is possible to construct the modified current:
\begin{equation}
\label{57}
J^{\mu}=\bar{\psi}\Gamma^{\mu}\psi
\end{equation}
in which the correction coefficients due to LIV, present in modified Gamma matrices and in the modified spinors, simplify, so this current is defined in the normal tangent space $(T_{x}M,\eta_{\mu\nu})$. Now using this result we can introduce the interaction term with the standard QED:
\begin{equation}
\label{58}
J^{\mu}A_{\mu}=J^{\mu}\eta_{\mu\nu}A^{\nu}=\bar{\psi}\Gamma^{\mu}\psi A_{\mu}=\bar{\psi}\Gamma^{\mu}\psi\eta_{\mu\nu} A^{\nu}
\end{equation}
We note that in this effective theory the interaction with the electromagnetic field takes places in the tangent space $(T_{x}M,\eta_{\mu\nu})$. It is significant to highlight that it is possible to obtain the same result introducing a Lorentz-violating extension to the Standard Model, as illustrated in \cite{Koste2}. In fact, considering only CPT even perturbation terms of the form:
\begin{equation}
\label{59}
\frac{1}{2}i\,c_{\mu\nu}\overline{\psi}\gamma^{\mu}\overleftrightarrow{D}^{\nu}\psi
\end{equation}
we can define the modified Dirac matrices in the form:
\begin{equation}
\label{60}
\Gamma^{\mu}=\gamma^{\mu}+c^{\mu\nu}\gamma_{\nu}
\end{equation}
demonstrating that, with an opportune choice of the terms violating the Lorentz invariance, it is possible to obtain analogue results to this work extending the Standard Model.\\
Even if more complicated than QED, interaction in the SM can be modified in the same way. Taking into account the modified Dirac equation, we can introduce for the quark sector a modified effective Lagrangian with the form:
\begin{equation}
\label{61}
\frac{i}{2}\sum_{j}\bar{\psi}_{j}\Gamma^{\mu}\overleftrightarrow{D}\psi_{j}
\end{equation}
where $D_{\mu}$ represents the flat covariant derivative of the SM, so even in this case spinorial and Cartan connections are negligible. Again, it is remarkable that in this case, the extension of the Standard Model proposed in \cite{Koste2} is compatible with the form obtained here. We can conclude that the introduced modified Standard Model lives in a asymptotically flat space-time. This result will be useful in the computation of important quantities introduced in the next section, such as the inelasticity.

\section{Effects induced on UHECR phenomenology}
Because of the presence of the Cosmic Microwave Background (CMB), Universe is not transparent for the propagation of high energy particles, which in fact interacts with the photons of the CMB and dissipate energy during their path. Owing to this effect, high energy particles are attenuated after a determined propagation length, which depends on their energy and their nature (type of the particle).  UHECR are constituted by heavy nuclei (iron type) or protons and the ways they interact with the background radiation are different. Heavy nuclei with sufficient energy suffer by CMB for example a photo-dissociation process:
\begin{equation}
\label{62}
A+\gamma\rightarrow (A-1)+n
\end{equation}
where $A$ represents the atomic number of the bare nucleus considered.\\
Instead protons interacts with the CMB only via a photo-pion production, through a delta resonance:
\begin{equation}
\label{63}
\begin{split}
&p+\gamma\rightarrow \Delta\rightarrow p+\pi^{0}\\
&p+\gamma\rightarrow \Delta\rightarrow n+\pi^{+}
\end{split}
\end{equation}
The last effect is known as GZK cut-off, from the name of three physicists, who first predicted this phenomenon (Greizen, Zatsepin, Kuzmin). In this work we will concentrate principally on the effects of LIV on the propagation of protons, so we will analyze this aspect.\\
In order to obtain a delta resonance the free energy of proton and photon must be bigger than the rest energy of the delta particle and this poses a constrain on the magnitude of LIV, in the approximation of head on collision:
\begin{equation}
\label{64}
\begin{split}
&s=(E_{p}+E_{\gamma})^2-(\overrightarrow{p}_{p}[e^{-1}]+\overrightarrow{p}_{\gamma})^2\geq m_{\Delta}^{2}\;\Rightarrow\\
\Rightarrow\;&(E_{p}+E_{\gamma})^2-\left(\frac{\overrightarrow{p}_{p}}{\sqrt{1-f_{p}(p_{p})}}+\overrightarrow{p}_{\gamma}\right)^2\geq m_{\Delta}^{2}\;\Rightarrow\\
\Rightarrow\;&E_{p}^{2}-\overrightarrow{p}_{p}(1-f_{p}(p_{p}))-2f_{p}(p_{p})\overrightarrow{p}_{p}+2E_{p}E_{\gamma}+\\
&-2\overrightarrow{p}_{p}\cdot\overrightarrow{p}_{\gamma}\left(1+\frac{1}{2}f_{p}(p_{p})\right)\geq m_{\Delta}^{2}\\
\end{split}
\end{equation}
where the four momentum of the proton is $(E_{p},\overrightarrow{p}_{p})$, the four momentum of a CMB photon is $(\omega,\overrightarrow{\omega})$, $m_{p}$  denotes the proton mass and $m_{\Delta}$ denotes the delta resonance mass. In the previous calculus it has been used MDR (\ref{11}) and the approximation:
\begin{equation}
\label{65}
\frac{1}{\sqrt{1-f_{p}(p_{p})}}\simeq 1+\frac{1}{2}f_{p}(p_{p})
\end{equation}
In the first line we have used the tetrad to project the momentum of the proton from its space of definition $(T_{x}M,\,g_{\mu\nu})$ to the space where the interaction between the massive lepton and the photon takes place, $(T_{x}M,\,\eta_{\mu\nu})$.\\
From (\ref{64}) we obtain the inequality:
\begin{equation}
\label{66}
2f_{p}(p_{p})E_{p}^{2}-E_{p}(4E_{\gamma}+f_{p}(p_{p})E_{\gamma})+m_{\Delta}^{2}-m_{p}^{2}\leq0
\end{equation}
If this second grade inequality is not satisfied, the GZK effect is suppressed as a consequence of LIV. From the study of this second grade inequality, we obtain, for the GZK existence, the constrain:
\begin{equation}
\label{67}
f_{p}(p_{p})<\frac{\Delta M^2-4E_{\gamma}-\sqrt{(\Delta M)^2-8E_{\gamma}\Delta M^2}}{4E_{\gamma}^2}
\end{equation}
where $\Delta M^2=(m_{\Delta}^{2}-m_{p}^{2})$.
Considering for these physical quantities the average values of $E_{\gamma}\simeq 7.0\times 10^{-4}\;eV$, $m_{\Delta}\simeq 1232\;MeV$, $m_{p}\simeq 938\;MeV$, we obtain a constrain:
\begin{equation}
\label{68}
f_{p}(p_{p})< 6\cdot10^{-23}
\end{equation}
to guarantee the existence of the GZK effect\footnote{This constrain is comparable to the superior limit $4.5\cdot10^{-23}$ obtained numerically in \cite{Scully2}}.\\
Protons lose energy by the photo-pion production process, and then they decay in protons or neutrons, without annihilate. So if they have enough energy, the process can repeat again, and it is necessary to evaluate the fraction of initial proton momentum transferred to the outgoing pion. To obtain this it is necessary to introduce the $\emph{elasticity}$ factor $\eta=\left(\frac{E_{out}}{E_{in}}\right)$, that is the ratio of the energy carried away by one of the particles emerging from the interaction, $(E_{out})$, divided by the energy of the incident particle, $(E_{in})$, and the \emph{inelasticity}, which represents the fraction of the total incident energy that is avaiable for the production of secondary particles, and is defined as $K=(1-\eta)$. Now if this phenomenon is not suppressed, it is possible to determine the \emph{attenuation length} or \emph{optical depth} of a proton, defined as the average length of propagation that a proton has to travel to see its energy reduced by a factor of $e^{-1}$ and is calculated integrating the probability of interaction of the Cosmic Microwave Background (CMB) with a proton propagating in the Universe:
\begin{equation}
\label{69}
\tau_{p\gamma}=\frac{1}{l_{p\gamma}}=\int_{E_{thr}}^{+\infty}dE\int_{-1}^{+1}d\mu\frac{1-\mu}{2}n(E)\sigma_{p\gamma}(s)K(s)
\end{equation}
where $\mu=\cos(\theta)$, $\sigma_{p\gamma}$ is the cross section for the photon-proton interaction, $K$ is the inelasticity and $n(E)$ is the distribution of the CMB, that is the Planck's formula for the energy dependent photon density in black body radiation:
\begin{equation}
\label{70}
n(E)=\frac{1}{\pi^2}\frac{E^2}{e^{E/KT}-1}
\end{equation}
Changing the perspective and considering the Mandelstam variable $s$ in the rest frame of the proton, where the photon four momentum is $(\omega',\,\overrightarrow{p}'_{\gamma})$, and using the fact that
\begin{equation}
\label{71}
\begin{split}
&s=(m_{p}+\omega')^2-\overrightarrow{p}_{\gamma}^2=m_{p}^2+2m_{p}\omega'\\
&\omega'=\gamma\omega(1-v_{p}\cos{\theta})\simeq2\omega\gamma\;\;\;\;(head\,on\,collision)
\end{split}
\end{equation}
the inverse of the mean free path becomes:
\begin{equation}
\label{72}
\frac{1}{l_{p\gamma}}=\tau_{p\gamma}=\int_{E_{thr}}^{+\infty}dE\,n(E)\int_{-1}^{+1}(1-v_{p}\cos{\theta})\frac{d\cos{\theta}}{2}\sigma_{p\gamma}(s)K(s)
\end{equation}
where $n(E)$ is the CMB energetic distribution and $v_{p}$ is the proton velocity.\\
In the case of UHECR (ultra high energy cosmic rays), that is protons with very high energy, it is possible to consider\footnote{Using the system of units of measurement for which the speed of light is equal to 1, this means taking the speed of protons equal approximately to that of light} $v_{p}\simeq 1$ and taking $ds=-2E_{p}\omega d\cos{\theta}$, the inverse of the mean free path becomes:
\begin{equation}
\label{73}
\tau_{p\gamma}=\frac{1}{8p^2}\int_{E_{thr}}^{+\infty}dE\,\frac{n(E)}{E^2}\int_{s_{min}}^{s_{Max}}ds\,s\sigma_{p\gamma}(s)K(s)
\end{equation}\\
where $E_{thr}$ is the threshold energy for the reaction and finally it is possible to obtain for the optical depth:
\begin{equation}
\label{74}
\tau_{p\gamma}=\frac{-KT}{2\pi^2\gamma^2}\int_{E_{thr}}^{+\infty}d\omega\,\sigma_{p\gamma}(\omega)K(\omega)\omega\ln(1-e^{-\omega/2KT\gamma})
\end{equation}
formula originally obtained by Stecker \cite{Stecker1,Harari1}.\\
Introducing the LIV, it is useful to notice that the previous computation must be conducted in a flat frame of reference, because the propagation of UHECR happens in an asymptotically flat space-time and the theory underlying the interaction between UHECR protons with CMB is the asymptotically flat SM minimal extension introduced before. So the most evident effect on the photo-pion process is limited to the modification of the inelasticity function, that is a modification of the allowed phase space for this kind of reaction. The inelasticity, calculated without introducing Lorentz violation in the theory, is given \cite{Stecker1} by the formula:
\begin{equation}
\label{75}
K(s)=\frac{1}{2}\left(1+\frac{m_{p}^2-m_{\pi}^2}{s}\right)
\end{equation}
Instead the modified dynamic, generated by the introduction of LIV, induces some changes in its computation. In fact, following the computation of \cite{Scully1,Scully2} and starting from the definition of the center of momenta reference frame:
\begin{equation}
\label{76}
\overrightarrow{p}_{p}^{*}+\overrightarrow{p}_{\pi}^{*}=0
\end{equation}
where these vectors are defined in $(TM,\,\eta_{ab})$, and considering the free energy of the photo-pion production:
\begin{equation}
\label{77}
\sqrt{s}=(E_{p}^{*}+E_{\pi}^{*})
\end{equation}
it is possible to obtain the $\gamma_{CM}$ factor, that allows to change frame of reference from the center of momenta to a generic one:
\begin{equation}
\label{78}
\gamma_{CM}(E_{p}^{*}+E_{\pi}^{*})=\gamma_{CM}\sqrt{s}=(E_{p}+E_{\pi})\;\Rightarrow\;\gamma_{CM}=\frac{E_{p}+E_{\pi}}{\sqrt{s}}=\frac{E_{tot}}{\sqrt{s}}
\end{equation}
Now evaluating the free energy necessary for the creation of a photo-pion in the CM frame of reference, and using the CM definition, so $\overrightarrow{p}^{*}_{p}=\overrightarrow{p}^{*}_{\pi}$:
\begin{equation}
\label{79}
\begin{split}
&(\sqrt{s}-E^{*}_{p})^2-([e_{\pi}]p_{p}^{*})^2=m_{\pi}^{2}\Rightarrow \\
\Rightarrow&(s-2\sqrt{s}E^{*}_{p})+E_{p}^{*2}-p_{p}^{*2}(1-f_{p})-p_{p}^{*2}f_{p}+p_{p}^{*2}f_{\pi}=m_{\pi}^{2}
\end{split}
\end{equation}
where $f_{p}$ and $f_{\pi}$ represent the LIV correction functions, introduced in eq.(\ref{9}), for the proton and the pion respectively.\\
From the previous relation follows:
\begin{equation}
\label{80}
E_{p}^{*}=\frac{s+m_{p}^2-f_{p}p_{p}^{*2}-m_{\pi}^{2}+f_{\pi}p_{\pi}^{*2}}{2\sqrt{s}}=F(s)
\end{equation}
and we can approximate:
\begin{equation}
\label{81}
\begin{split}
&p^{*}_{p}\simeq E'_{p}=(1-k_{\pi}(\theta))\sqrt{s}\\
&p^{*}_{\pi}\simeq E'_{\pi}=k_{\pi}(\theta)\sqrt{s}
\end{split}
\end{equation}
where we use the final energies, those of the final products after the interaction, and $\sqrt{s}$ represents the initial free total energy.\\
From the change of reference frame and approximating the coordinate change equations with the Lorentz invariant ones, it is possible to write:
\begin{equation}
\label{82}
E_{p}=\gamma_{CM}(E_{p}^{*}+\beta \cos{\theta} p_{p})
\end{equation}
where $E_{p}=(1-k_{\pi}(\theta))E_{tot}$, using the pion inelasticity.\\
Substituting $\gamma_{CM}$ with the value computed in eq. (\ref{68}) and approximating the three-momentum magnitude with the energy and the velocity factor $\beta$ with 1, in the hypothesis of ultra-relativistic particles, we obtain the following equation:
\begin{equation}
\label{83}
(1-k_{\pi}(\theta) )=\frac{1}{\sqrt{s}}\left(F(s)+\cos{\theta}\sqrt{F(s)^2-m_{p}^2+2f_{p}}\right)
\end{equation}
from which it is possible to obtain the inelasticity in function of the collision angle $\theta$. The quantity must then be averaged on the interval $\theta\in[0,\;\pi]$ to obtain the inelasticity used in the computation:
\begin{equation}
\label{84}
k_{\pi}=\frac{1}{\pi}\int_{0}^{\pi}k_{\pi}(\theta)\;d\theta
\end{equation}

\begin{figure}
  \centering
  \includegraphics[width=0.75\textwidth]{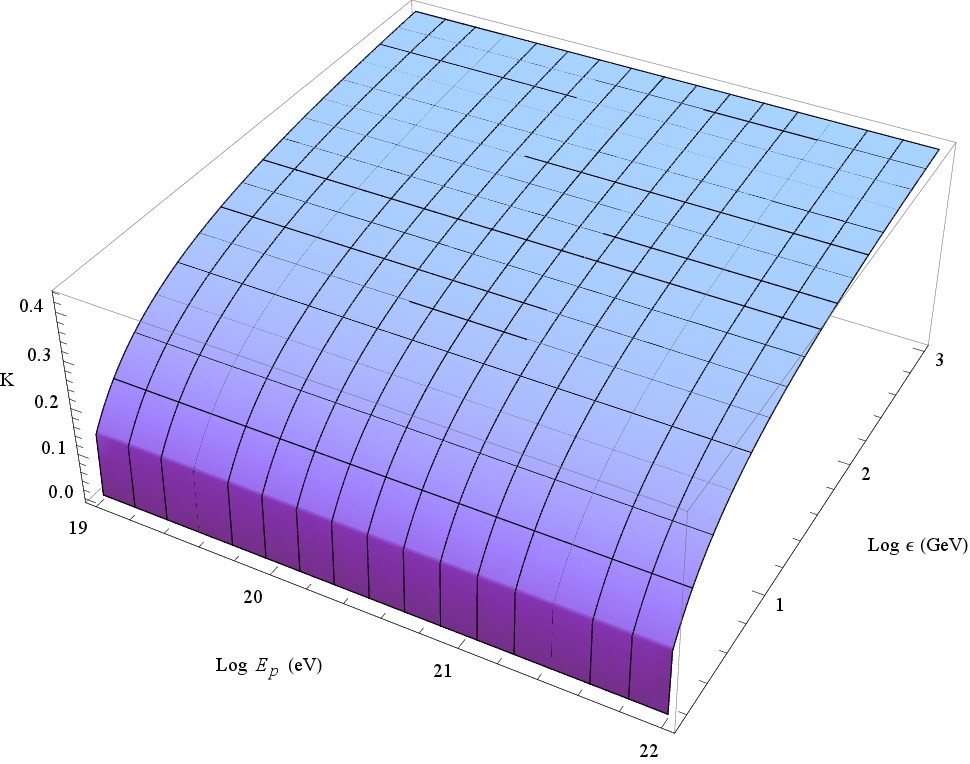}
  \includegraphics[width=0.75\textwidth]{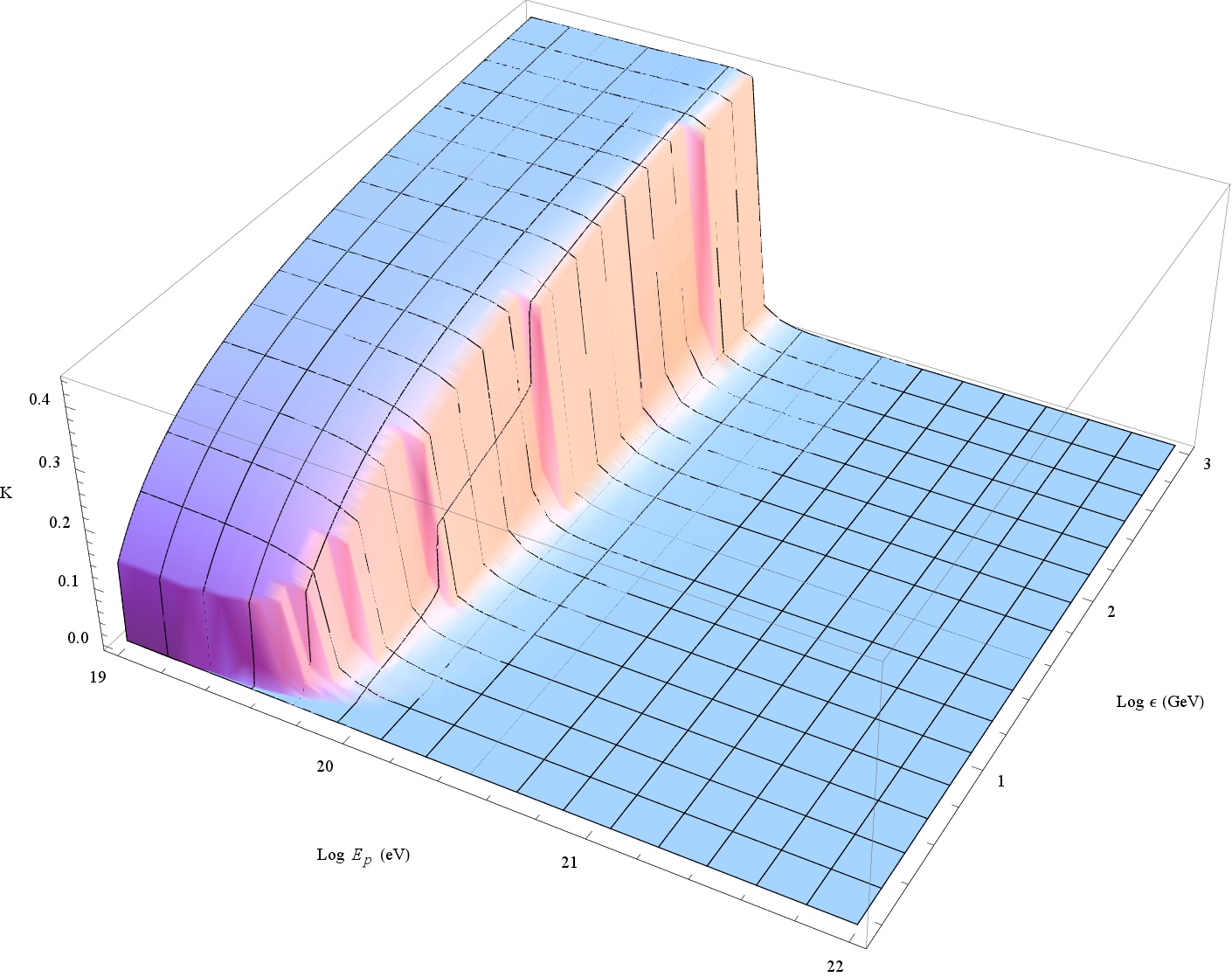}
  \caption{Respectively inelasticity for perturbation $f_{p\pi}\simeq 9\cdot10^{-23}$ and inelasticity for $f_{p\pi}=0$}\label{fig:def1}
\end{figure}
\begin{figure}
  \centering
  \includegraphics[width=1\textwidth]{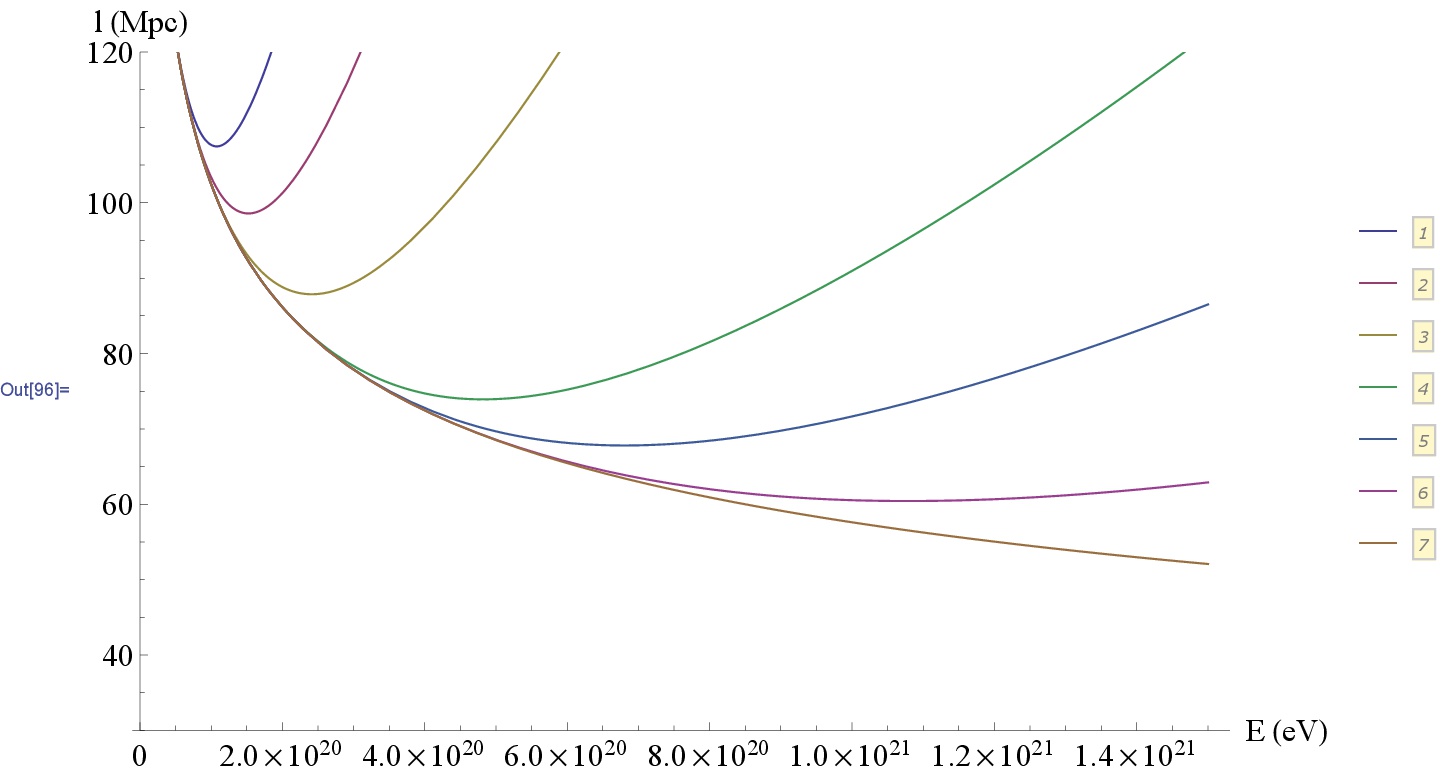}
  \includegraphics[width=1\textwidth]{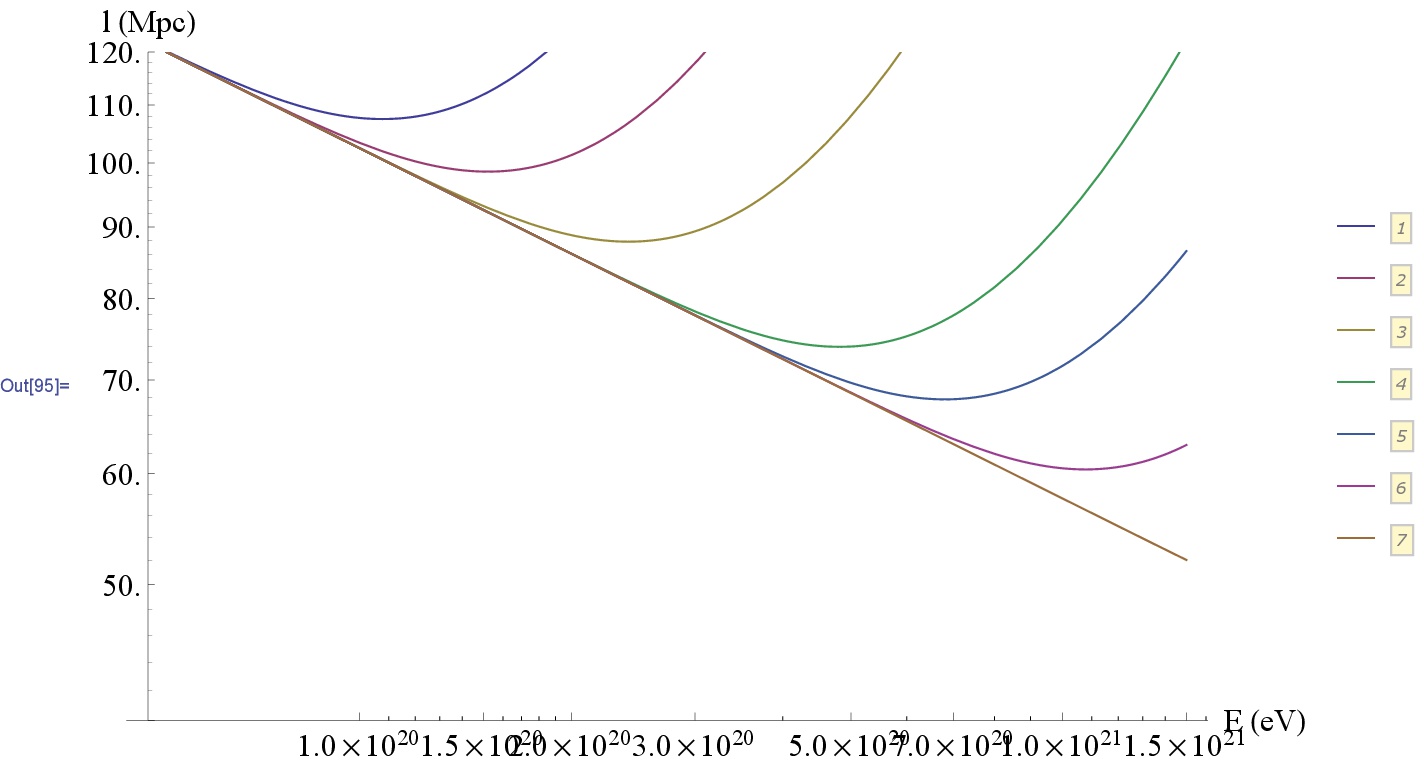}
  \caption{Optical depth as function of energy and LIV parameters, respectively: $1)\,f_{p\pi}\simeq 9\cdot10^{-23}$, $2)\,f_{p\pi}\simeq 6\cdot10^{-23}$, $3)\,f_{p\pi}\simeq 3\cdot10^{-23}$, $4)\,f_{p\pi}\simeq 9\cdot10^{-24}$, $5)\,f_{p\pi}\simeq 6\cdot10^{-24}$, $6)\,f_{p\pi}\simeq 3\cdot10^{-24}$, $7)\,f_{p\pi}\simeq 0$  } \label{fig:def}
\end{figure}

In fig. \ref{fig:def1} we illustrate an example of the effects of a tiny perturbation, inferior to constrain (\ref{68}), due to the introduction of LIV, on the value of the inelasticity of the photo-pion process, expected to become $1/2$ for high energies. The consequent modification of the expected optical depth of a proton as a function of its energy and of the LIV magnitude, is plotted in fig. \ref{fig:def}. We obtained that this plot depends on the difference between the magnitude of the perturbation correlated with the proton and the pion: $f_{p\pi}=f_{p}-f_{\pi}$. In this work we consider only LIV perturbations which imply that every particle has a maximum attainable velocity lower than $c$ and it is physically reasonable to expect the more massive one having smaller velocity, that is a bigger violation, which corresponds to $f_{p}>f_{\pi}$. It is important to underline that even for very tiny violations of Lorentz Invariance, the effects can be absolutely relevant, implying a consistent dilatation of the predicted GZK opacity sphere.

\section{Closing Remarks}
As already underlined in literature \cite{Biete1}, cosmic rays constitute a fundamental source of information, because they involve the most energetic particles in the Universe and they propagates on extremely long paths, so they are the ideal candidates to probe the quantum structure of space-time. In this work we have explored the effects of this quantum structure on the propagation of Ultra High Energy Cosmic Rays, introducing Lorentz Invariance Violation. A first consequence is the necessity to modify the dispersion relations governing the dynamics of particles. We have focused on dispersion relations modified by homogeneous perturbations, because this approach permits a continuous transition from the GZK phenomenon to the Coleman and Glashow foreseen total suppression. Furthermore in this way we preserve the MDRs origin from metrics. So we have given a theoretical background, in the form of an effective theory, to the Modified Dispersion Relations, and we have illustrated the possibility to resort to Finsler geometry to explain the dynamics \cite{Koste1}. Moreover we have shown that even tiny Lorentz Invariance perturbations can have dramatic effects on UHECR. In fact we have estimated the increase of the optical depth for protons, due to a reduction of their maximum attainable velocity. In conclusion we have shown that LIV can modify the expected GZK opacity sphere, so investigations on Universe transparency to cosmic rays can give interesting results on the validity of such departures from standard physics.

\section*{Acknowledgements}
We would like to thank prof. Alan Kostelecky and prof. Stefano Liberati, for reading a draft of this paper, suggesting some relevant improvements, prof. Sean T. Scully, for the help with the Mathematica code for determining the inelasticity. We would also like to thank Marco Gherardi, for having shared his computer science skills with us.

\appendix

\section{Legendre transformation in a Finsler space}
Here we report the proof of the proposition exposed in section 2 where is introduced the concept of Finsler geometry.\\
\emph{Proposition}:
\begin{enumerate}
  \item $F=F^{*}\circ l$
  \item \emph{the Legendre transformation is a bijection}.
\end{enumerate}
\emph{Proof}: \\
The first Legendre transformation property can be demonstrate considering the fact that:
\begin{equation}
\label{84}
F(v)=\frac{g{y}_{ij}y^{i}y^{j}}{F(y)}=l_{y}\left(\frac{y}{F(y)}\right)\leq F^{*}\circ l(y)
\end{equation}
and a symmetric inequality, given by:
\begin{equation}
\label{85}
F^{*}\circ l(y)=\sup_{v\neq0}{l_{y}\left(\frac{v}{F(v)}\right)}=\sup_{v\neq0}{\left(\frac{g(y)_{ij}y^{i}v^{j}}{F(v)}\right)}\leq F(y)
\end{equation}
The injectivity of the Legendre transformation follows from:
\begin{equation}
\label{86}
g(y)_{ij}y^{i}w{j}=g(v)_{ij}v^{i}w^{j}\quad \forall w\in V \;\Rightarrow v=y
\end{equation}
in fact posing $w=v$ and $w=y$ we have:
\begin{equation}
\label{87}
\begin{split}
&F^2(v,v)=g(v)_{ij}v^{i}v^{j}=g(y)_{ij}y^{i}v^{j}\leq F(v)F(y)\\
&F^2(y,y)=g(y)_{ij}y^{i}y^{j}=g(v)_{ij}v^{i}y^{j}\leq F(y)F(v)
\end{split}
\end{equation}
where the Cauchy-Schwarz\footnote{It can be proved that the Cauchy-Schwarz inequality is still valid even in the case of Finsler geometry.} inequality has been used.\\
The surjectivity can be proved observing that, if $\xi\in V^{*}\backslash0$, $\lambda=F^{*}(\xi)$ and $y\in V$ such that $F(y)=1$ and $\xi(y)=\lambda$. Now considering the smooth curve $\gamma:I\rightarrow F^{-1}(1)$:
\begin{equation}
\label{88}
\gamma(t)=\frac{y+tw}{F(y+tw)},\quad t\in I
\end{equation}
where the vector $w$ is such that $w\in\{w\in V:g(y)_{ij}y^{i}w^{j}=0\}$, because $y$ is the maximum of the function $v\rightarrow \xi(v)$, it follows that:
\begin{equation}
\label{89}
0=\frac{d}{dt}\xi(\gamma(t))\mid_{t=0}=\xi\left(\frac{w}{F(y)}-\frac{y}{F^{2}(y)}\frac{\partial F}{\partial y^{i}}(y)w^{i}\right)
\end{equation}
so, because $g(y)_{ij}y^{i}w^{j}=0$, it follows that $\xi(w)=0$. From this, one sees that $\forall v\in V$ the following decomposition is possible:
\begin{equation}
\label{90}
v=g(y)_{ij}y^{i}v^{j}+w
\end{equation}
From these results follows that $\xi=l(\lambda y)$, proving the surjectivity.

\section{Modified Lorentz group and homgeneity of correction functions}
The form of the transformations of the modified Lorentz group defined in section 5 are:
\begin{equation}
\label{91}
\Lambda^{\mu}_{\;\nu}=\Lambda^{a}_{\;b}[e]_{a}^{\;\mu}[e]^{b}_{\;\nu}
\end{equation}
Acting on the 4-vector $p^{\mu}=(E,\,\overrightarrow{p})$, this gives, for the modification function $f$ in the MDR:
\begin{equation}
\label{92}
f\left(\frac{|\overrightarrow{p}|}{E}\right)\rightarrow f\left(\frac{|\Lambda^{i}_{\;\mu}(f(p))p^{\mu}|}{\Lambda^{0}_{\;\mu}(f(p))p^{\mu}}\right)
\end{equation}
and it is simple to verify that this kind of transformations preserve the homogeneity of degree 0, because of the ratio present in the definition of the modification function $f$.

\section{Lorentz violating extension of the Standard Model}
In this work we have considered MDRs that are equal for particles and antiparticles, because we deal only with protons. This corresponds to modify the Standard Model introducing only CPT-even therms, as illustrated in \cite{Koste2}. Furthermore the MDRs form selected does not distinguish between particle polarizations. In fact, considering a SM extension with CPT even terms of the form:
\begin{equation}
\label{93}
\frac{1}{2}i\,c_{\mu\nu}\overline{\psi}\gamma^{\mu}\overleftrightarrow{D}^{\nu}\psi+\frac{1}{2}i\,d_{\mu\nu}\overline{\psi}\gamma_{5}\gamma^{\mu}\overleftrightarrow{D}^{\nu}\psi
\end{equation}
it is possible to define the modified Dirac matrices:
\begin{equation}
\label{94}
\Gamma^{\mu}=\gamma^{\mu}+c^{\mu\nu}\gamma_{\nu}+d^{\mu\nu}\gamma_{5}\gamma_{\nu}
\end{equation}
obtaining an effective Lagrangian that induces an MDR with a difference, taking into account that one is dealing with real fermions (particles with spin). Again, supposing this difference smaller than other terms involved in modifying the SM, we can neglect this tiny difference, in order to obtain the effects of interaction of UHECR with CMB.

\end{document}